\documentclass[12pt]{article}

\usepackage{graphicx}
\usepackage{scicite}
\usepackage{amsmath}

\usepackage{times}

\usepackage{amsmath}
\usepackage{color}
\usepackage{soul}

\newenvironment{sciabstract}{%
\begin{quote} \bf}
{\end{quote}}

\newcounter{lastnote}

\usepackage[T1]{fontenc}
\usepackage[utf8]{inputenc}
\usepackage{authblk}

\title{More than one Author with different Affiliations}
\author[1]{Haiming Deng}
\author[1]{Lukas Zhao}
\author[2]{Kyungwha Park}
\author[3]{Jiaqiang Yan}
\author[4]{Kamil Sobczak}
\author[1,5$\dagger$]{Lia Krusin-Elbaum}

\small
\affil[1]{Department of Physics, The City College of New York - CUNY, New York 10031, USA }
\affil[2]{Department of Physics, Virginia Tech, Blacksburg, Virginia 24061, USA}
\affil[3]{Materials Science and Technology Division, Oak Ridge National Laboratory, Oak Ridge, Tennessee 37831, USA}
\affil[4]{Faculty of Chemistry, Biological and Chemical Research Center, University of Warsaw, 02-089 Warsaw, Poland}
\affil[5]{City University of New York Graduate Center, New York, New York 10016, USA}
\date{}

\normalsize
\title{Topological surface currents accessed through reversible hydrogenation of the three-dimensional bulk} 

\begin{document}


\baselineskip20pt


\maketitle
\vspace{-8mm}


\begin{sciabstract}
Hydrogen, the smallest and most abundant element in nature, {can be efficiently  incorporated within a solid
and drastically modify its electronic state --- it has been known to induce novel magnetoelectric effects in complex perovskites}
and modulate insulator-to-metal transition in a correlated Mott oxide.
Here we demonstrate that hydrogenation resolves an outstanding challenge in chalcogenide classes of three-dimensional (3D) topological insulators and magnets 
--- the control of intrinsic bulk conduction {that denies access to quantum surface transport}.
With electrons donated by a reversible binding of H$^+$ ions to Te(Se) chalcogens, carrier densities are easily {changed}
by over $10^{20}~\textrm{cm}^{-3}$, {allowing tuning the Fermi level into the bulk bandgap to
enter surface/edge current channels}.
{The hydrogen-tuned topological materials are stable at room temperature and tunable
disregarding bulk size, opening
a breadth of platforms for harnessing emergent topological states.}
\end{sciabstract}


\newpage

The ability to control carrier density --- a key parameter of the electronic state of condensed matter ---  is central to achieving access to the topologically protected surface states in three-dimensional (3D) topological materials (TIs) that ideally are insulating in the bulk \cite{Qi2011}.  In the absence of magnetic dopants, the 2D electronic surface states with Dirac-type (linear) energy-momentum dispersion \cite{Qi2011} are gapless and fully spin-polarized, with protection against backscattering by local disorder guaranteed by time-reversal symmetry. These important layered van der Waals (vdW) quantum materials have narrow ($\simeq 300$ meV) bulk gaps \cite{Zhang-NatPhys09} and charge carriers donated by intrinsic lattice defects \cite{Scanlon-antisites2012}.
{As a result, bulk conduction can dominate 
the surface channels, hindering
the studies of the Dirac states and their implementation in topological spintronics
and fault-tolerant quantum computing \cite{spintronics-Qcomp2019}.}

{Recent realizations of a nontrivial quantum anomalous Hall (QAH) state, featuring dissipation-free chiral edge currents \cite{QAHReview-SCZhang2016}, made it apparent that
when long-range magnetism and band topology combine \cite{MagTop-Tokura2019}
the problem of bulk conduction can be particularly acute. Magnetic dopants, by breaking time-reversal symmetry, gap out the Dirac surface channels \cite{ChenMassiveDirac2010}, and when the Fermi level is in the Dirac (mass) gap the QAH state is expected to emerge. However, in addition to magnetic moment, the dopants also donate charge.
Indeed, QAH was first observed in heavily Cr-doped (Bi,Sb)$_2$Te$_3$ ultrathin films \cite{CZC-QAH1st_exp2013} in which to minimize bulk conduction a non-stoichiometric alloying of the constituent elements 
was employed and the film thickness had to be precisely controlled. Moreover, doping disorder
restricted QAH temperature to the mK range.}

{In a newly discovered important class of intrinsic topological magnets (ITM) \cite{surface-states-SL-2019} the doping and alloying disorders are avoided because the magnetic dopants are arranged as an atomic layer in a layered van der Waals (vdW) crystal structure. For example, the Mn-based ITM MnBi$_2$Te$_4$ consists of Te-Bi-Te-Mn-Te-Bi-Te septuple layers (SLs), separated by vdW gaps and coupled antiferromagnetically  \cite{Otrokov2019}. As a consequence, the anomalous Hall (near) quantization was observed at higher temperatures \cite{QAH-MBT-Science2020}. However, such materials show high bulk carrier densities (typically $> 10^{20}~ \textrm{cm}^{-3}$)\cite{MBST-evolution2019,Sb-MBT-phases2019}, and observation of QAH required delicate tuning of charge density that was again only possible in ultrathin (3-5 nm) samples with odd number of SLs.}
In thicker (hundreds of nanometers) samples  ferromagnetism and quantization could be achieved when SLs were separated by a topological spacer \cite{QAH-Haiming2020},
yet tuning to quantization of either ferromagnetic (FM) or antiferromagnetic (AFM) ITMs without imposing severe thickness constraints is a major obstacle to realizing novel topological quantum states.

Here we demonstrate a remarkably efficient and facile way to reversibly tune bulk carrier densities by over $\simeq 10^{20}~\textrm{cm}^{-3}$ {in different classes of} layered topological materials by using insertion and extraction of ionic hydrogen to achieve access to topological surface states. The source of H$^\textrm{+}$ ions is a dilute aqueous hydrochloric acid (HCl) solution,  which leaves the TI crystal structure
intact and has an extra benefit of removing native surface oxide while passivating surfaces and preventing reoxidation under exposure to air. We show that H$^{+}$ in TIs, {by preferentially forming an H-Te(Se) moiety within van der Waals gaps}, donates electrons to the system and moves Fermi level from the bulk-valence (BVB) to bulk-conduction (BCB) bands, {crossing the bulk bandgap to display an ambipolar \textit{p}- to \textit{n-}type conductance conversion at the charge neutrality point.}
{The process is fully reversible, as hydrogen-chalcogen moiety can be disassociated by a low-temperature annealing protocol under which
hydrogen is easily removed. It is also multiply-cyclable and reproducible, thereby resolving one of the key limitations of
magnetic and nonmagnetic TIs.}

{Generally, crystal growth of tetradymite crystals such as Bi$_2$Te$_3$ and Bi$_2$Se$_3$ results in equilibrium defect configurations comprising vacancies and antisites on different sublattices \cite{Scanlon-antisites2012}. 
In undoped Bi$_2$Se$_3$, where Se vacancies {tend} to dominate, the net conduction is by electron carriers or \textit{n}-type.  In Bi$_2$Te$_3$, where antisites are prevalent \cite{Drasar2010} the conductivity is usually \textit{p}-type, namely by hole carriers, although by varying stoichiometry it has been grown of either conductivity type.}
{Exploring growth and the exact doping/alloying conditions for the Fermi level $E_F$ to lie within the narrow bulk gap is a lengthy and often chancy process.} By contrast, we show that a post-growth hydrogenation can tune $E_F$ to {charge neutrality}
in hours, and can be effective in single crystals and in
nano-devices, where the evolution of resistance under hydrogenation is easily monitored over time.

{The ultralight hydrogen diffuses rapidly and
easily incorporates into many materials \cite{H-storage-matls-2001}, although it shows qualitatively different behavior in different hosts \cite{H-and-defects-1992}.}
It has been shown to modify both electronic and structural states \cite{H-VO2-2-2016,H-perovskite3-2014,GR-graphane-2009,H-perovskite2-2017} --- turning  graphene into graphane \cite{GR-graphane-2009}, inducing new magnetic phases in complex perovskites \cite{H-perovskite2-2017} and modulating insulator-to-metal transition in a correlated Mott oxide VO$_2$ \cite{H-electolyte_gating_VO2-4-2016}.
{In most semiconductors interstitial hydrogen binds to defects \cite{H-and-defects-1992}
and is known to be amphoteric \cite{HinSemicons-2006}, namely it can act either as a donor (H$^+$) or an acceptor (H$^-$) of charge, nearly always counteracting
the prevailing conductivity type.}
To ascertain the effects of hydrogen uptake on charge transport in the topological materials we first chose
well characterized Bi$_2$Te$_3$ crystals with the conductivity flavor initially of a net acceptor type. The level of hydrogenation was controlled by timing the diffusion of H$^+$ ions from a dilute (0.5 molar) HCl + H$_2$O = H$^\textrm{+}$(H$_2$O) + Cl$^\textrm{-}$ solution maintained at room temperature in which the samples were immersed for a period of minutes to hours (see {Figs. 1A,B and Materials and Methods}).

Our key and a most immediately notable result shown in {Figure 1} is {a nearly two orders of magnitude} increase in the low-temperature longitudinal resistance $R_{xx}$ as a function of HCl exposure time to a maximum $R_{xx}^{\textrm{max}}$ and a subsequent decrease as the exposure time is prolonged {(Fig. 1C)}. The resistance maximum $R_{xx}^{\textrm{max}}$ (conductance minimum) is at the charge-neutral point (CNP) where conduction is converted from \textit{p-} to \textit{n-}type, as determined from the corresponding Hall resistance ({Fig. 1D}). We surmise then that hydrogen, through the process of in-diffusion into the bulk {(see Supplementary Note 1)}, indeed donates electrons. This process is not self-limiting: at long ($\geq 24~\textrm{h}$) exposure times a decrease in $R_{xx}$ on the deep \textit{n}-side {slows down yet continues}.
The observed ambipolar conduction, with well-distinguished \textit{p} (hole) and \textit{n} (electron) conduction regions,
can be reproducibly reversed when hydrogen is removed by a low-temperature anneal ({sketch in Fig. 1B}) in the easily accessible range that depends on sample thickness.
For the $\sim 200~\textrm{nm}$ thick Bi$_2$Te$_3$ crystal shown here the de-hydrogenation and a return from \textit{n-}type across the CNP back to \textit{p-}type is {obtained within $\sim 100^o\textrm{C}$}. {Importantly, the carrier mobility  during the entire process is not affected --- e.g. in the sample in {Fig. 1} it remains at $\cong 7,000~ \textrm{cm}^2 \textrm{V}^{-1} \textrm{s}^{-1}$
{(Supplementary Table 1)}.}

After each annealing step the system is room-temperature stable and can be further finetuned by the conventional electrostatic gating. Figure 1E shows that once the carrier density is sufficiently reduced by H$^\textrm{+}$, the vicinity of the CNP can be reproducibly explored by applying a gate voltage $V_g$. $R_{xx}(V_g)$ is precisely reversible upon changing the direction of the $V_g$ sweep --- the absence of hysteresis is a clear indication that hydrogenation does not create surface traps \cite{Evangelos-PRM2018} and is consistent with hydrogen permeating the bulk.
{We emphasize that in thick ($\geq 100~\textrm{nm}$) materials, far away from the CNP where carrier densities are in the high $\sim 10^{19}~ \textrm{cm}^{-3}$ range,
the CNP can not be reached by gating {within a practically accessible voltage sweep ({Fig. 1F}).}

The type conversion on hydrogenation and de-hydrogenation is clearly seen in the magnetic field dependence of Hall resistance $R_{xy}$ ({Fig. 2A}), with $R_{xy}$ flipping its slope $\textrm{d}R_{xy}/\textrm{d}H$ and Hall coefficient $R_H = -\frac{1}{n_b e}$ changing sign in the conversion region. {Figure 2B} shows a characteristic low-field ambipolar behavior of $R_{xy}$ near the CNP, where the net residual bulk carrier density is very low. {As we approach the CNP, weak antilocalization (WAL) quantum correction to `classical' conductivity emerges at low magnetic fields as a \textit{positive} magnetoresistance cusp ({Fig.~ 2C}),  characteristic of a TI \cite{irrad-Lukas2016}. The number $n_Q$ of quantum conduction channels contributing to WAL can be estimated from 2D localization theory \cite{HLN1980}
$\Delta G(B)  \simeq  \alpha \frac{e^{2}}{2\pi ^{2}\hbar}f(\frac{B_{\phi }}{B})$,
where $\Delta G(B)$ is the low-field quantum correction to 2D magnetoconductance, coefficient $\alpha = n_Q/2$ equals to 1/2 for a single 2D channel, $f(x) \equiv ln x -\psi(1/2 + x)$, $\psi$  is the digamma function, and field $B_{\phi }=\frac{\hbar}{4el_{\phi}^{2}}$ is related to the dephasing length  $l_{\phi }$ of interfering electron paths. {The fit to WAL conductance {(see Supplementary Fig. 1)} yields $\alpha \simeq 1.004 \pm0.001$, corresponding to two 2D quantum channels we associate with top and bottom topological surfaces \cite{irrad-Lukas2016}.} {Figure 2D} schematically depicts the Bi$_2$Te$_3$ bandstructure \cite{Zhang-NatPhys09} as the Fermi level $E_F$ moves from BVB to BCB (hydrogen in) and in reverse
(hydrogen out). {The 2D WAL cusp is seen when $E_F$ is within the Dirac bands}.}

{The change in the net carrier density induced by hydrogenation is reflected in the Shubnikov-de Haas (SdH) quantum oscillations of Hall resistance at higher fields ({Fig. 2B}). An estimate of the Fermi surface size obtained from the change of the oscillation period in $\delta R_{xy}/\delta \mu_0 H$ ({Fig. 2E-G}) shows that the
Fermi vector $k_F$ is much reduced
after hydrogenation {(Supplementary Table 2)} --- {near the CNP {$k_F \approx 0.014${\AA}$^{-1}$} and the corresponding carrier density
{$n \cong\times 10^{16}\textrm{cm}^{-3}$} is very low.}
After a {longer time exposure} (5.8 h) to HCl, as $E_F$ enters BCB the Fermi surface size becomes large again, albeit a more complex one (with three oscillation periods, {see Fig. 2, Supplementary Fig. 2}), reflecting the Bi$_2$Te$_3$ bandstructure with \textit{n-}type Dirac bands mixing with BVB}.

Next we ask, where does hydrogen go? Our transport experiments demonstrate that in Bi$_2$Te$_3$ {hydrogen donates electrons and stabilizes} in its positive charge state as H$^{+}$, a proton. To optimize its Coulomb interaction with the host electrons, it will tend to locate in the regions of higher electron charge density, i.e. in the vicinity of a more electronegative Te.
To chase its physical presence we first utilize time-of-flight secondary ion mass spectrometry (ToF-SIMS),
focusing on Te {(Figs. 3A-D, Supplementary {Fig. 3})}. A negative bias mass spectrum of secondary ions ({see Materials and Methods)} ejected from Bi$_2$Te$_3$  shows a cluster of peaks in the 120-130 {u (atomic mass)}
range ({Fig. 3C}) that belong to several naturally occurring isotopes of Te,
see {Supplementary Table 3}. A closer look at the Te isotope cluster ({Fig.~3D}) clearly shows additional peaks that differ from each Te isotope by {exactly} 
{$\Delta \textrm{u}=1$}, consistent with each isotope binding with one hydrogen.

On a microscopic scale, we confirm the affinity between H and Te by x-ray photoelectron spectroscopy (XPS), see Methods.
XPS spectra near the Bi 4f  and Te 3d  core levels show that, despite a strong doping effect of hydrogen, the Bi core levels \cite{XPS-Bi2Te3-2013} are unperturbed ({Fig. 3E}), while the spectral shapes of Te 3d peaks are quantifiably modified {(Fig. 3F,G)}. The upshift $\Delta = 1~\textrm{eV}$ in binding energy due to [Te-H]$^-$ moiety is easily distinguished from free hydrogen or surface oxidation, as seen, for example, during  hydrogen electro-catalysis \cite{H-catalysisBi2Te3-2020}, where Bi 4f spectral shapes are upshifted by $\sim 2~\textrm{eV}$ and both  Bi$_2$O$_3$  and TeO$_2$ peaks are present. {By contrast, XPS shows that surface oxide is cleanly removed by HCl, and that Cl
is entirely absent {(Supplementary Fig. 4 and Table 4)}}.

The electron-donor action of hydrogen in a vdW TI through formation of hydrogen-chalcogen moiety is borne out by the density functional theory (DFT) calculations (see {Supplementary Note 2} and {Supplementary Figs. 5-7}). The calculated band structures of Bi$_2$Te$_3$ ({Fig. 4A-C}) show that after hydrogenation {the Dirac bands are preserved while the Fermi level $E_F$ is upshifted into the BCB}. This \textit{n}-type doping is independent of whether H$^+$ goes in interstitially or into the van der Waals gaps, however the H-Te moiety appears most stable within the vdW gap (the formation energies are in line with XPS). The calculated local density of states (DOS) ({Fig. 4D}) {highlights the shift of $E_F$ relative to the Dirac point (DP)}.
Bi$_2$Te$_3$ structure maintains charge balance within each quintuple layer (QL) of atoms Te$^{(1)}$-Bi--Te$^{(2)}$-Bi-Te$^{(1)}$, where the superscripts distinguish the inequivalent Te sites {(see Supplementary Fig. 6a)}.
The Te-Bi bonds are polar-covalent while only weak vdW type bonding exists between the neighboring Te$^{(1)}$--Te$^{(1)}$ planes with relatively weak and highly polarizable bonds. The hydrogen prefers going to the region between these planes {(Fig. S7)}, while the crystal structure remains intact {(Supplementary Fig. 8)}.

The Fermi level tuning by hydrogenation demonstrated here is very general, indeed it is remarkably effective in other chalcogen-based TI. We have achieved the type conversion across the CNP in Se-containing materials, such as Bi$_2$Te$_2$Se ({Supplementary Fig. 9}) and Ca-doped Bi$_2$Se$_3$ {(Supplementary Figs. 10,11)}. Our DFT calculations confirm that a stable [H-Se]$^- $ moiety similarly acts to move the Fermi level {towards the conduction} bands ({Supplementary Figs. 6,7}).

The technique is also very powerful in magnetic TI, and particularly in a very promising ITM class.
One example is the intrinsic ferromagnetic MnBi$_2$Te$_4$/Bi$_2$Te$_3$ superlattice \cite{QAH-Haiming2020} where initial bulk carrier density is in the $10^{20}~ \textrm{cm}^{-3}$ range, yet type conversion across the CNP by hydrogen is obtained  {(Supplementary Figs. 12,13)}. {Hydrogenation can also deliver an essentially continuous $E_F$ tuning across the bulk gap of an intrinsic topological antiferromagnet, where the axion insulator behavior was very recently reported when the antiferromagnetic ITM was made thin enough ($\sim 6$ SLs or $\sim 8$ nm) \cite{Axion-Chern_Insulators-AFM2020} to reduce the contribution from bulk conduction channels and to be able to deplete bulk carriers by electrostatic gating. We demonstrate that through hydrogen-tuning of $E_F$ we can turn the bulk AFM ITM {into an insulating state} sans voltage gating. {Figure 4E} shows how the longitudinal resistance $R_{xx}(T)$ of a ten times thicker MnBi$_{2-x}$Sb$_x$Te$_4$ ($x = 0.6$) crystal continuously transforms under hydrogenation from a metallic-like into a strongly insulating-like  ---  a feat not achieved by a step-wise Bi-Sb alloying \cite{MBST-evolution2019}. Under hydrogen uptake the N\'{e}el temperature $T_N \sim 25~K$ remains unchanged while the low-temperature $R_{xx}(T)$ increases by about two orders of magnitude, reaching maximum value at the CNP ({Fig. 4F}). {The Hall resistance $R_{xy}$ ({Fig. 4G}) at the CNP displays ambipolar behavior, akin to the one in  Bi$_2$Te$_3$ ({Fig. 1D})}.
The field dependencies of $R_{xx}(H)$ and $R_{xy}(H)$ under hydrogenation are consistent with the results in thin flakes under voltage gating \cite{QAH-MBT-Science2020,Axion-Chern_Insulators-AFM2020}, {also see} {Supplementary Fig. 14}.
{Near} the CNP the magnetoresistance $R_{xx}(H)$ sharply changes at two characteristic {fields}: $H_1$, above which the spin order is driven into a canted AFM state \cite{MBST-evolution2019,Sb-MBT-phases2019}, and $H_2$
at which all spins align in a FM state ({Fig. 4H}).
Below $H_1$,
Hall resistance $R_{xy}(H)$ exhibits a distinct plateau 
(where some of the SLs align with field) and a `zero-plateau' ({Fig. 4I}). {Together with the rapidly decreasing $R_{xx}(H_1<H<H_2 )$ these are the key signatures of surface currents in close proximity to the Chern \cite{Top-theoryQi2008} and axion  \cite{AxionI-SL-2019} insulator states.}

Finally, we remark that a paucity of bulk-insulating topological materials significantly impedes the search for emergent topological quantum phenomena, with the prospect for real-world applications remaining far off. Hydrogen-tunability of high bulk carrier densities expands the availability of robust and easily accessible platforms for observing and harnessing distinct topological phases with stunning macroscopic manifestations, such as dissipationless edge transport of charge and axion electrodynamics \cite{Top-theoryQi2008} in topological magnets. It should facilitate the search for quantized magnetoelectric \cite{MagElectric-zero-plateauQAH2015} and magnetooptical  \cite{Qu-Topmagoptics2020} effects, and could lead to higher-temperature  
and higher-order
topological states
for future antiferromagnetic spintronics \cite{AFMspintronics2018} and quantum computing. }


\small
\bibliography{refsMain262,booksetc10}
\bibliographystyle{Science}



\normalsize
\vspace{2mm}
\noindent \textbf{Acknowledgements:} We wish to thank Tai-De Li for the technical assistance with the surface analytical tools
at the Surface Science Facility of CUNY Advanced Science Research Center (ASRC). {Jim Hone's probing comments are much appreciated.} This work was supported by the NSF grants DMR-1420634, DMR-2011738, and HRD-1547830.  Computational support was provided by Virginia Tech ARC and San Diego Supercomputer Center (SDSC) under DMR-060009N.

\vspace{2mm}
\noindent\textbf{Author Contributions} Experiments were designed by H.D. and L.K.-E.. {Materials were grown by H.D. and J.Y.}, and their modification and characterization, device fabrication and transport measurements were performed by H.D. and L.Z.. {HAADF-STEM and EDX elemental imaging was performed by K.S..} DFT bandstructures and electronic density of states (DOS) under hydrogenation were calculated by K.P.. Data analysis was done by H.D. and L.K.-E. L.K.-E. wrote the manuscript with the input from H.D.
$^{\dagger}$Correspondence should be addressed to L. K.-E (email: lkrusin@ccny.cuny.edu).
\vspace{3mm}



\vspace{-5mm}
\noindent\section*{FIGURE LEGENDS}
\noindent \textbf{Fig.~1.Tuning bulk conductivity of a topological material by hydrogenation.}
\textbf{A,B}, Illustration of hydrogenation and de-hydrogenation process.
\textbf{A}, {Hydrogenation: sample is submerged in} a dilute aqueous HCl solution (0.5M) at room temperature
where H$^+$ permeates the bulk through a timed diffusion process.
\textbf{B}, {De-hydrogenation: H$^+$ inside the sample} is controllably released
as H$_2$ gas by an anneal in the $70-200^\circ\textrm{C}$ temperature range prescribed by the sample thickness.
\textbf{C}, Longitudinal resistance $R_{xx}$ and
{\textbf{D}, Hall resistance $R_{xy}$ at 2 K and $- 5.7$ T of the initially \textit{p-}type Bi$_2$Te$_3$ 120nm-thick exfoliated device
{\textit{vs.} hydrogenation time (bottom axis) show conversion (following blue arrows) to \textit{n}-type (blue symbols). The back-conversion (red symbols) to \textit{p}-type (following red arrows) is controlled by thermal annealing steps (30 min at each temperature, top red axis).}}
$R_{xx}$ increases by nearly two orders of magnitude at the $\sim$ CNP. The maximum is at the ambipolar point in $R_{xy}$.
{\textit{Top inset: }Bi$_2$Te$_3$ bandstructure rendering. \textit{Bottom inset}: {Optical image of the sample with van der Pauw contact geometry used.}}
{\textbf{E}, Electrostatic back-gating in the vicinity of the CNP (labeled \textbf{1}) reveals the expected maximum in $R_{xx}$ at the CNP.}
The non-hysteretic $R_{xx}$ \textit{vs}. gating voltage $V_g$ indicates an absence of surface traps.
{\textbf{F}, Far from the CNP on the \textit{n}-type (labeled \textbf{2}) or
\textit{p}-type side (labeled \textbf{3}),} the $R_{xx}$ maximum is inaccessible by voltage gating.

\vspace{2mm}

\noindent \textbf{Fig.~2. Transport and quantum oscillations across charge neutrality point.}
\textbf{A}, Hall resistance $R_{xy}$ \textit{vs}. magnetic field $H$ after hydrogenation and on annealing at different temperatures $T_a$. Initially, the pristine Bi$_2$Te$_3$ crystal is \textit{p}-type. The conversion from \textit{p- }to \textit{n-}type and back is indicated by the sign change of the slope $dR_{xy}/dH$.
\textbf{B}, In the compensated region near the CNP there are both carrier types from the surfaces and the bulk. At {higher} fields Hall slope is negative {and SdH oscillations are observed}, as expected from the \textit{n}-type Dirac surface states (see Fig. 2\textbf{D}).
{\textit{Inset}: nonlinear Hall resistance near zero field.}
\textbf{C}, Evolution of magnetoresistance {(normalized to the value at zero field)} under annealing with time steps $\Delta t = 30~ \textrm{min}$ implemented to tune Bi$_2$Te$_3$ crystal to stable CNP; it evolves form a quadratic field dependence of a typical bulk metal to a weak antilocalization (WAL) regime with a characteristic low-field cusp near CNP. {The cusp fit parameter $\alpha = 2 \times 0.5$ signifies contributions from top and bottom surfaces (0.5 from each), see text and Supplementary Fig. 1.}
\textbf{D}, Bandstructure cartoon for Bi$_2$Te$_3$ illustrates the upshift of the Fermi level $E_F$ on hydrogenation (cyan arrow) and the reversal by de-hydrogenation (red arrow). In Bi$_2$Te$_3$, the $E_F^0$ at the CNP is slightly above the Dirac point (DP) \cite{Qi2011,Zhang-NatPhys09}.
Shubnikov-de Haas (SdH) quantum oscillations of $\delta R_{xy}/\mu_0\delta H$ in Bi$_2$Te$_3$ with magnetic field applied along the \textit{c}-axis:
\textbf{E}, before hydrogenation,
\textbf{F}, near the CNP and
\textbf{G}, after 5.8~h exposure to H$^+$. The corresponding sizes of Fermi surfaces
are drawn as circles.

\vspace{2mm}

\noindent \textbf{Fig.~3. Spectroscopic detection of hydrogen in Bi$_2$Te$_3$.}
\textbf{A}, Time-of-flight secondary ion mass spectrometry (ToF-SIMS) generates mass spectrum of the outermost 1.5-2.0 nm surface region by bombarding it
with a primary ion (Bi$^+$) beam.
{\textbf{B}, ToF-SIMS images of Te (\textit{left}), Bi (\textit{middle}), and Si  substrate (\textit{right}) indicate the rastered area from which the spectra were obtained (see Methods).  {The scale bar is 10 $\mu$m.}  
 \textbf{C}, {Mass analysis of the secondary ions under negative bias shows the presence of $^1$H, $^{128}$Te, $^{209}$Bi, and {$^{129}\textrm{[Te-H]}$} moieties.} Well separated peaks of the known long-lived Te isotopes with the intensity ratios occurring in nature {(Supplementary Table 3)} are detected.
{\textbf{D}, The expanded view of the Te isotope {atomic mass} range clearly shows additional peaks (highlighted in blue) of lower intensity that perfectly and sequentially follow each  Te$^\alpha$ peak by precisely $\Delta u =1$. Some hydrogen peaks appear to overlap with Te$^\alpha$.}
{\textbf{E}, XPS data of Bi 4f$_{7/2}$ and  4f$_{5/2}$ for pristine Bi$_2$Te$_3$ and H$^+$ treated {for $t = 12$ hours} show no change after hydrogenation.
\textbf{F}, Te 3d$_{3/2}$ and  3d$_{6/2}$ are modified by hydrogenation.}
\textbf{G}, Analysis of the Te {3d} XPS peaks clearly shows H binding to Te with a shift in binding energy of  $\Delta = 1 eV$.

\vspace{2mm}

{\noindent \textbf{Fig.~4. DFT-calculated bandstructures under hydrogenation and hydrogen-induced conductivity type conversion in a magnetic TI.}
DFT bandstructures for four quintuple layer (QL) slabs of
\textbf{A}, pristine Bi$_2$Te$_3$,
\textbf{B}, with interstitial hydrogen
(H-Te$^{(2)}$) bonding) within QLs and
\textbf{C}, with hydrogen in the vdW gap (H-Te$^{(1)}$ bonding). The red color indicates the states localized at the top or bottom surface states.
{\textbf{D}, Calculated total electronic densities of states (DOS). \textit{Top}: For pristine Bi$_2$Te$_3$.
\textit{Middle}: With hydrogen in the vdW gaps acting as electron donor.
\textit{Bottom}: Cl would act as electron acceptor, moving $E_F$ in the opposite direction --- the doping action not observed experimentally in transport and consistent with the absence of Cl in XPS.}
{\textbf{E}, Longitudinal resistance $R_{xx}(T)$ of a $\sim 80~\textrm{nm}$ thick MnBi$_{2-x}$Sb$_x$Te$_4$ ($x = 0.6$) crystal clearly shows hydrogen-induced transformation from a metallic to an insulator-like state. Inset: The characteristic cusp (red arrow) occurs at the N\'{e}el temperature $T_N \sim 25~K$, which is not affected by hydrogenation.
\textbf{F}, Longitudinal resistance $R_{xx}$ and
\textbf{G}, Hall resistance $R_{xy}$ of the same crystal at 1.9 K (below $T_N$), both as a function of hydrogenation time and annealing temperature $T_a$. The maximum of $R_{xx}$ is at the CNP where the conduction is converted from \textit{p-} to \textit{n-}type, and correspondingly the ambipolar behavior is observed in $R_{xy}$.}
{\textbf{H}, $R_{xx}$ and
\textbf{I}, Hall resistance $R_{xy}$ at 1.9 K measured at the CNP as a function of applied magnetic field. Below $T_N$ and above the characteristic field $H_1\approx 3~\textrm{T}$ (green arrows) the system is driven into a canted AFM state, and becomes fully FM aligned at $H_2\approx 6~\textrm{T}$ (black arrows).
{The surface Hall currents switch from a zero $R_{xy}$ plateau to a finite plateau when some of the SLs align with the field.}
The $R_{xy}$ plateaus observed below $H_1$ are hallmark signatures of the Chern and axion insulator states.
}

\vspace{5mm}

\newpage

\hspace{-20mm}
\includegraphics[width=1.17\textwidth]{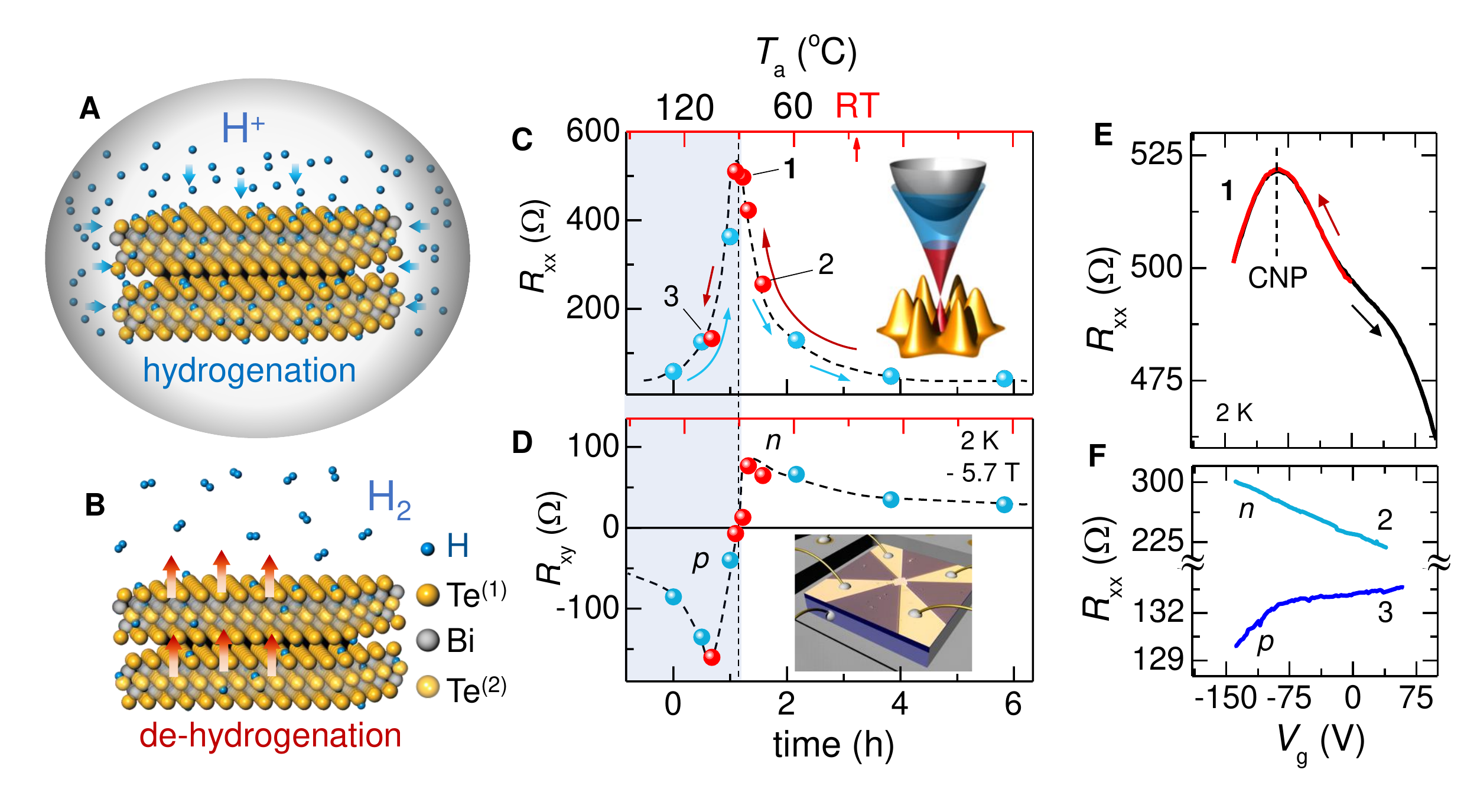}
\vfill\hfill Fig.~1; {HD \it et al.} \eject

\hspace{-20mm}
\includegraphics[width=1.12\textwidth]{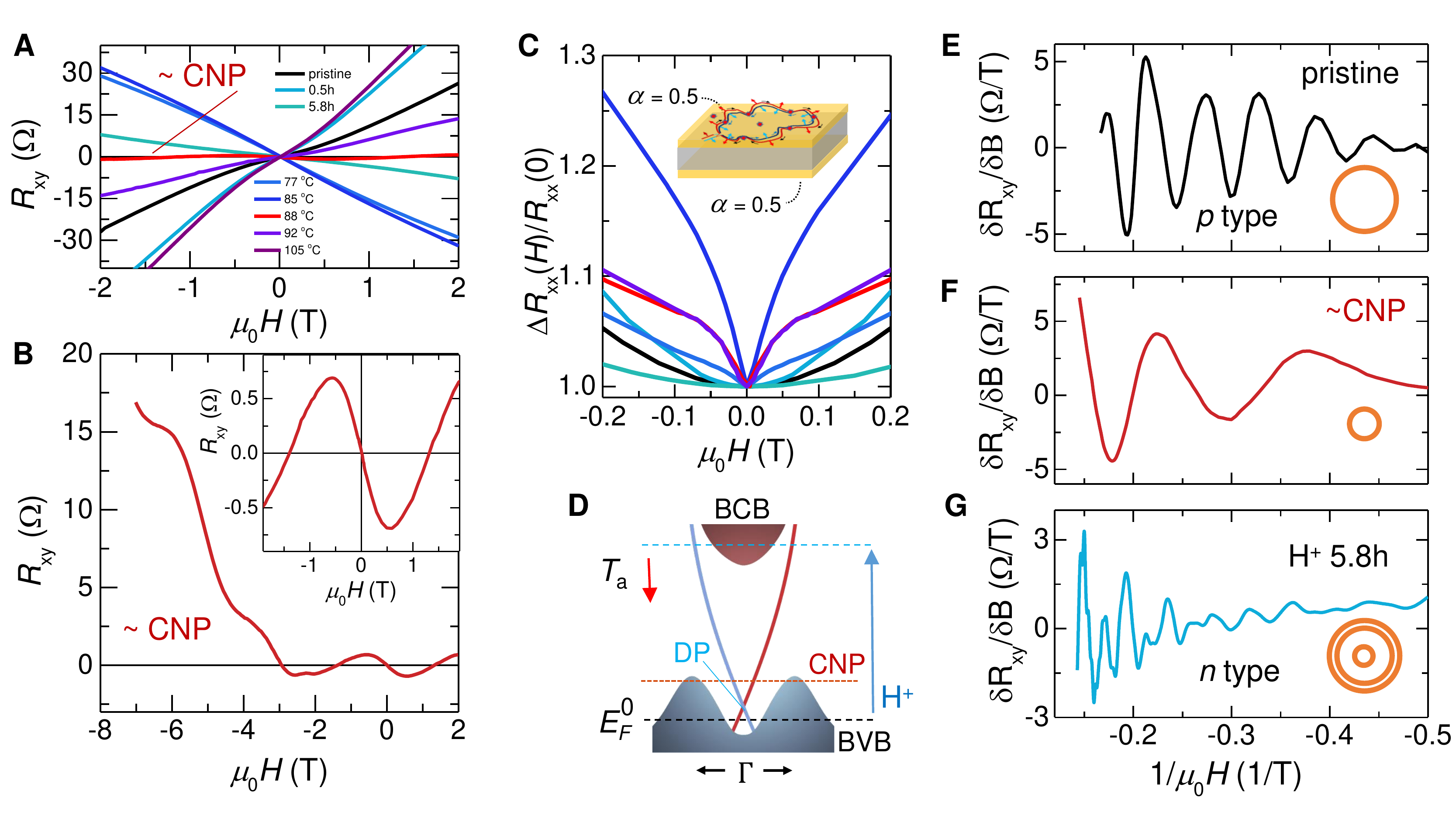}
\vfill\hfill Fig.~2; {HD \it et al.} \eject

\hspace{-18mm}
\includegraphics[width=1.1\textwidth]{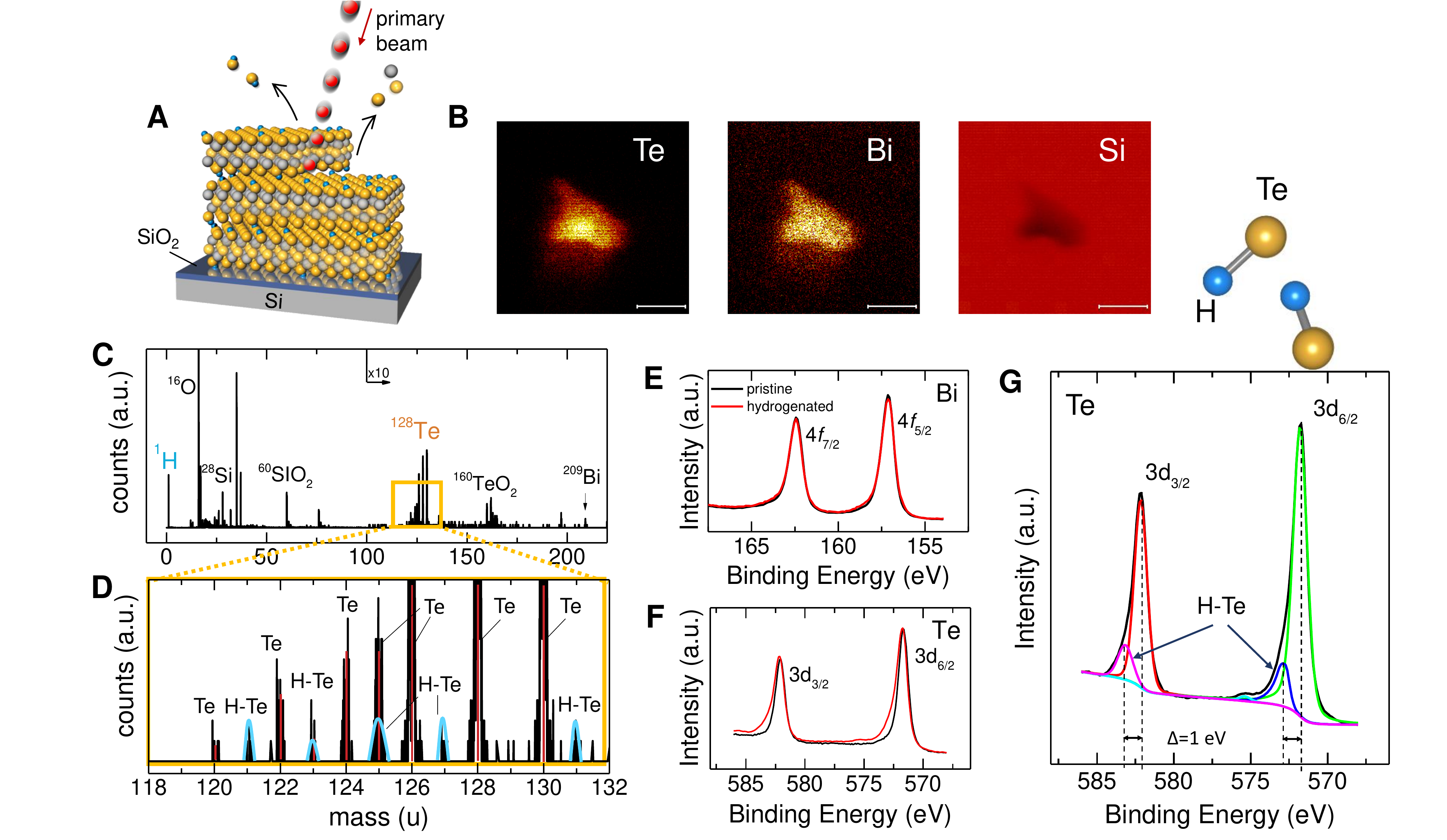}
\vfill\hfill Fig.~3; {HD \it et al.} \eject

\hspace{-20mm}
\includegraphics[width=1.1\textwidth]{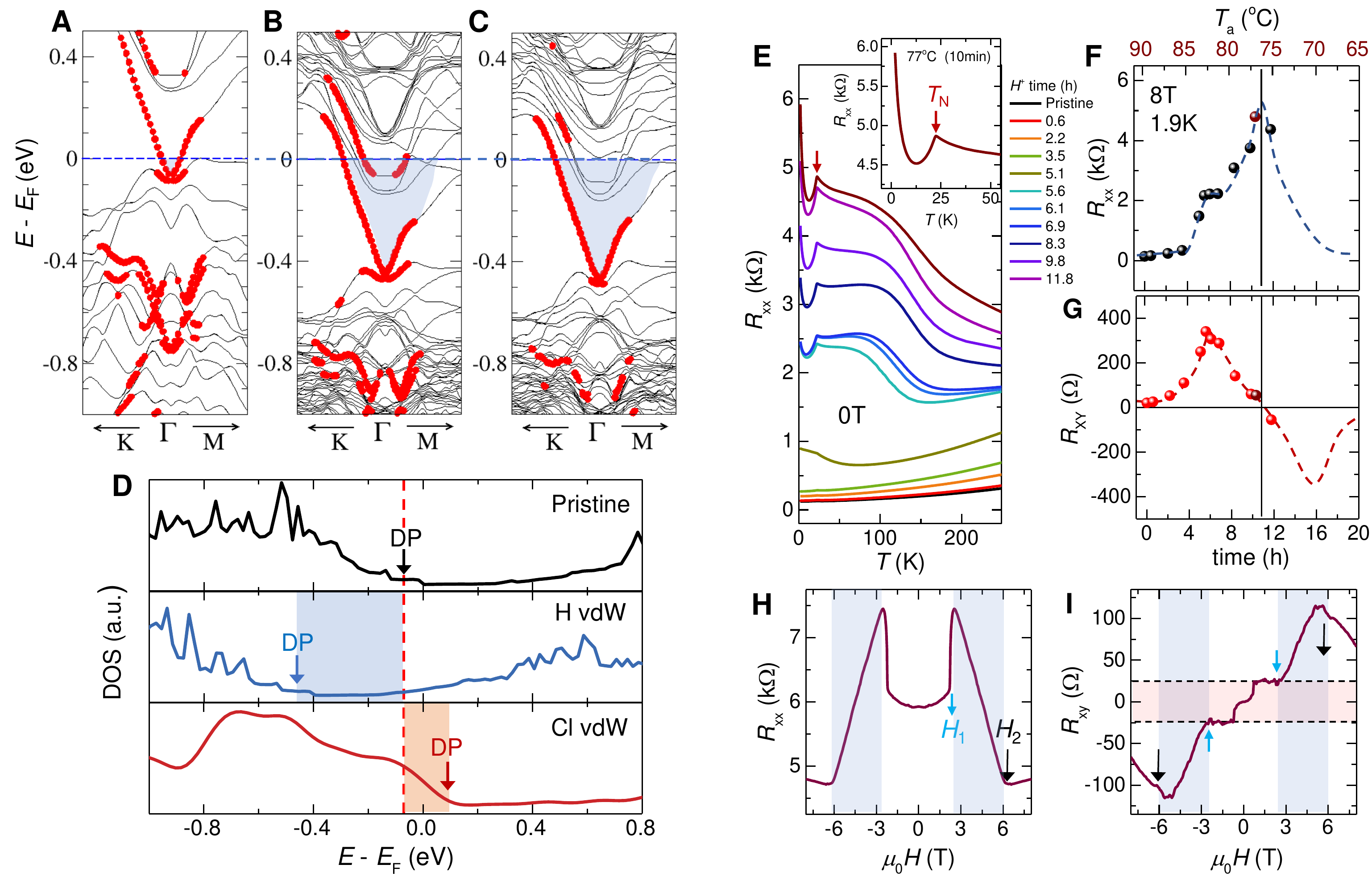}
\vfill\hfill Fig.~4; {HD \it et al.} \eject

\end{document}